\documentclass[floats,preprint,prb,aps]{revtex4}
\usepackage{graphicx}
\begin{document}


\date{\today}
\vspace{2.7in}

\title{Superfluidity of ``dirty'' indirect excitons and
magnetoexcitons  in two-dimensional trap }

\author{Oleg L. Berman$^{1}$, Yurii E. Lozovik$^{2}$,  David W.
Snoke$^{1}$, and Rob D. Coalson$^{3}$}

\affiliation{\mbox{$^{1}$Department of Physics and Astronomy,
University of
Pittsburgh,}  \\ 3941 O'Hara Street, Pittsburgh, PA 15260 USA   \\
\mbox{$^{2}$ Institute of Spectroscopy, Russian Academy of
Sciences,}  \\ 142190 Troitsk, Moscow Region, Russia  \\
\mbox{$^{3}$ Department of Chemistry, University of Pittsburgh,}
\\ Pittsburgh, PA 15260, USA}

\date{\today}

\vspace{2.7in}

\begin{abstract}
The superfluid phase transition
of bosons in a two-dimensional (2D)  system with disorder and
an external parabolic potential is studied. The theory is applied to
experiments on indirect excitons in coupled quantum wells. The random
field is allowed to be large compared to the dipole-dipole repulsion between
excitons. The slope of the external parabolic trap is assumed to
change slowly enough
to apply the local density approximation  (LDA) for the superfluid
density, which
allows us to calculate the Kosterlitz-Thouless temperature
$T_{c}(n(r))$ at each local
point $r$ of the trap. The superfluid phase occurs around the center
of the trap  ($\mathbf{r}=0$) with the normal phase outside this
area. As temperature increases, the superfluid area shrinks and
disappears at temperature $T_{c}(n(r=0))$.  Disorder acts to deplete the
condensate; the minimal total number of excitons for which superfluidity exists
increases with disorder at fixed temperature. If the disorder is
large enough, it can
destroy the superfluid entirely. The effect of magnetic field is also
calculated for
the case of indirect excitons. In a strong magnetic field $H$, the superfluid
component decreases, primarily due to the change of the exciton effective mass.

\vspace{0.2 in}



\end{abstract}

\maketitle {}


\section{Introduction}

Superfluidity in a system of spatially indirect excitons (with
spatially separated electrons and holes) in coupled quantum wells
(CQW) has been predicted by Lozovik and Yudson,\cite{Lozovik} and
several subsequent theoretical
studies\cite{Klyuchnik,Shevchenko,Lerner,Dzjubenko,Kallin,Knox,
Yoshioka,Birman,Littlewood,Vignale,Berman,Berman_Tsvetus,Berman_Willander,Ulloa}
have suggested that this should be
manifested as persistent electric currents, quasi-Josephson
phenomena and unusual properties in strong magnetic fields. In the
past ten years, a number of experimental studies have focused on
observing these
behaviors.
\cite{Snoke_paper,Snoke-ssc,Chemla,Krivolapchuk,Timofeev,Zrenner,Sivan,Snoke}
  One of the appeals of this system is that the electron
and hole wavefunctions have very little overlap, so that the
excitons can have very long lifetime  ($> 20$ $\mu$s), and
therefore they can be treated as metastable particles to which
quasiequlibrium statistics apply. Also, when the spatial
separation between the electrons and holes is large enough, the
interactions between the excitons are entirely dipole-dipole
repulsive.

In real experiments disorder plays a very important role.
\cite{Snoke_paper,Snoke-ssc,Chemla,Krivolapchuk,Timofeev,Zrenner,Sivan}
Although the inhomogeneous broadening linewidth of typical
GaAs-based samples has been improved from around 20 meV to less
than 1 meV,\cite{Snoke_paper} the disorder energy is still not
always small compared to the exciton-exciton repulsion energy. At
a typical exciton density of a few times $10^{10}$ cm$^{-2}$, the
interaction energy of the excitons is several meV.  On the other
hand, the typical disorder energy of 1 meV is low compared to the
typical exciton binding energy of 5 meV, so that the excitons can
be viewed as stable particles.  Typical thermal energies at liquid
helium temperatures are $k_BT = 0.2 - 2$ meV. Previous
work\cite{Berman_Snoke_Coalson} has shown that in the
low-temperature limit, in the case of no trapping potential, the
density $n_{s}$ of the superfluid component in CQW systems and the
temperature of the superfluid transition (the Kosterlitz-Thouless
temperature $T_{c}$\cite{Kosterlitz}) decrease with increasing
amplitude of the random field (the general description of
``dirty'' boson problem is presented in
Ref.~[\onlinecite{Vinokur}]).

In the present paper we consider the phase transition in the
system of the excitons trapped in a confining potential, in which
case true Bose condensation is possible. Experimentally, an in-plane harmonic
potential is possible in GaAs quantum well structures by using inhomogeneous
stress.\cite{Snoke-apl,Snoke-ssc} The Bose condensation in this case
is similar to
that for Bose atoms in
trap.\cite{Kim,Anderson,Ensher,Ketterle,Ketterle_Miesner,Daflovo,Pitaevskii}
  The slope of the external parabolic trap is assumed to change
slowly enough to apply the local density approximation  (LDA) for
the superfluid density and Kosterlitz-Thouless temperature at each
local point of the trap. This assumption allows us to substitute
the exciton density profile in the Thomas-Fermi
approximation\cite{Pitaevskii} into the expressions for the
superfluid density and Kosterlitz-Thouless temperature
obtained for the system without the confinement, found in
Ref.~[\onlinecite{Berman_Snoke_Coalson}]. We also analyze the
phase transitions in strong magnetic field, applying the effective
Hamiltonian approach developed in
Ref.~[\onlinecite{Berman_Snoke_Coalson_magnet}].

The paper is organized in the following way.  In Sec.~\ref{nomag}
the local density profile, superfluid density and
Kosterltz-Thouless transition in the trapped indirect exciton
system with disorder are obtained. The temperature dependence of the minimal
total number of excitons for superfluidity  at various levels of disorder is
calculated. In Sec.~\ref{mag} the local density profile and the phase
transitions of
the trapped dirty system in magnetic field are obtained. In
Sec.~\ref{discussion} we present our conclusions .


\section{The density profile and the local Kosterlitz-Thouless phase
transition in LDA}
\label{nomag}

We consider a slowly changing parabolic potential with the
characteristic length of inhomogeneity $l$ much greater than the
excitonic  de-Broglie wavelength $\lambda = \hbar/\sqrt{2M\mu}$:
\begin{eqnarray}
\label{Thomas_Fermi_cond} l \gg \frac{\hbar}{\sqrt{2M\mu}} ,
\end{eqnarray}
where $M = m_{e} + m_{h}$ is the mass of indirect exciton ($m_{e}$
and $m_{h}$ are the electron and hole masses, respectively), and
$\mu$ is the chemical potential of the system with  dipole-dipole
repulsion, given in the ladder approximation by\cite{Yudson} (here
and below, $\hbar = 1$)
\begin{equation}\label{Mu}
\mu =  \frac{8\pi n}{2M \log \left(\displaystyle \frac{\epsilon^{2}}{8\pi s^2 n
M^2 e^4 D^4} \right)} ,
\end{equation}
where $e$ is the charge of an electron, $D$ is the distance
between quantum wells, $s$ is the spin degeneracy, $\epsilon$ is
the dielectric constant, $n$ is the two-dimensional (2D) exciton
density without confinement (we consider the dilute exciton gas
$n\rho^{2} \ll 1$, where $\rho$ is the exciton radius).

Following earlier work,\cite{Ruvinsky_jetp} one can write the random field felt
by an indirect exciton in coupled quantum wells, induced by
fluctuations in the widths of electron and hole quantum wells, in the form
  $ V(\mathbf{r}_{e},\mathbf{r}_{h}) =
\alpha_{e}[\xi_{1}(\mathbf{r}_{e}) - \xi_{2}(\mathbf{r}_{e})] +
\alpha_{h}[\xi_{3}(\mathbf{r}_{h}) - \xi_{4}(\mathbf{r}_{h})]$,
where separate parameters are introduced for the random fields
felt by the electrons and holes which make up the exciton, since
in the standard coupled quantum well structure the electron and
hole are spatially separated in two different wells. Here
$\alpha_{e,h} = \partial E_{e,h}^{(0)}/\partial d_{e,h}$, where
$d_{e,h}$ is the average width of the electron or hole quantum
well, and $E_{e,h}^{(0)}$ is the lowest energy of the electron and
hole confined state; $\xi_{1}$, $\xi_{2}$, $\xi_{3}$ and $\xi_{4}$
are the fluctuation amplitudes of the upper and lower interfaces
of the two quantum wells. We assume that fluctuations on different
interfaces are statistically independent, whereas fluctuations of
a specific interface are characterized by Gaussian correlation
function $ \left\langle
\xi_{i}(\mathbf{r}_{1})\xi_{j}(\mathbf{r}_{2}) \right\rangle =
g_{i}\delta_{ij}\delta(\mathbf{r}_{2} - \mathbf{r}_{1})$, $
\left\langle \xi_{i}(\mathbf{r}) \right\rangle = 0$, where $g_{i}$
is proportional to the squared amplitude of the $i$th interface
fluctuation.\cite{Ruvinsky_jetp}

As shown in earlier work,\cite{Berman_Snoke_Coalson} in the low
frequency and long wavelength limits, the effect of this type of
disorder is to give a single parameter $Q$ which is an imaginary
self-energy of the exciton due to the disorder. The Green's
function of the center of mass of the isolated exciton at $T = 0$
in the random field is given in the momentum-frequency domain
within the second order Born approximation by\cite{Gevorkyan}
\begin{eqnarray}
\label{green_0} G^{(0)}(\mathbf{p}, \omega) = \frac{1}{\omega -
\varepsilon_{0}(p) + \mu + i Q(\mathbf{p}, \omega) }  ,
\end{eqnarray}
where $\mu$ is the chemical potential of the system,
$\varepsilon_{0}(p) = p^{2}/2M$ is the spectrum of the center mass
of the exciton in the ``clean'' system, and the parameter $Q$ is
determined by the disorder as
\begin{eqnarray}
\label{Q_res_ap}
Q (\mathbf{p}, \omega) &=& Q =
\frac{\alpha_{e}^{2}(g_{1} + g_{2}) -
\alpha_{h}^{2}(g_{3} +
g_{4})}{16\pi^{4}}M ; \hspace{0.5in}  m_{e} =
m_{h} ; \ \ \alpha_{h} \left(
\frac{m_{h}}{m_{e}}
\right)^{2} \ll \alpha_{e},
\nonumber \\
 Q (\mathbf{p}, \omega) &=&
Q =
\frac{\alpha_{e}^{2}(g_{1} + g_{2})}{64\pi^{4}}M  ,
\hspace{0.5in}
   m_{h} \gg m_{e}; \ \ \alpha_{h} \left(
\frac{m_{h}}{m_{e}}
\right)^{2} \ll
\alpha_{e}.
\end{eqnarray}

For the slowly changing confinement potential $U(r) = \gamma
 r^{2}/2$ ($\gamma$ is the curvature of the confinement potential applied to the
  center of mass of the exciton taken from Ref.[\onlinecite{Snoke-apl}],
   and $r$ is the distance between the center of mass of the
   exciton and the center of confinement), assuming the inequality
Eq.~(\ref{Thomas_Fermi_cond}) holds, the Thomas-Fermi
approximation can be applied in the thermodynamic
equilibrium.\cite{Pitaevskii} In the Thomas-Fermi approximation,
the chemical potential of the system at each local point equals
the sum of the coordinate-dependent local chemical potential and
the external field. Since for the ``dirty" system the chemical
potential enters to the interacting Green's function as
$\sqrt{\mu^{2} - Q^{2}}$ (which represents the chemical potential
of the ``dirty" infinite [unconfined] excitonic
system\cite{Berman_Snoke_Coalson}), we obtain the Thomas-Fermi
equation for the local density profile $n(r) = n_{0} - \Delta
n(r)$, (where $n_{0} = n(r = 0)$ is the exciton density at the
point of the minimum of the parabolic potential)
\begin{eqnarray}
\label{den_pr} \sqrt{\mu^{2}[n_{0}] - Q^{2}} =
\sqrt{\mu^{2}[(n_{0} - \Delta n(r))] - Q^{2}} + U(r),
\end{eqnarray}
where $\mu[n(r)]$ represents the contribution to the chemical
potential of the ``clean'' system due to dipole-dipole repulsion
at each local point, as found by substituting the
coordinate-dependent local density $n(r)$ into Eq.~(\ref{Mu}). In
the very dilute limit ($n_{0}\rho^{2} \rightarrow 0$, $\rho$ being
the exciton radius) for the weak confinement potential ($\Delta
n(r) \ll n_{0}$ and $\mu^{2}[n_{0}]\Delta n(r)/[(\mu^{2}[n_{0}] -
Q^{2})n_{0}]  \ll 1$) we get the approximate exciton density
profile
\begin{eqnarray}
\label{den_pr_ap} \Delta n(r) = \frac{\gamma}{2}
\frac{\sqrt{\mu^{2}[n_{0}] - Q^{2}}}{\mu^{2}[n_{0}]} n_{0} r^{2}.
\end{eqnarray}
The exciton density profiles at different $n_{0}$ and $Q$ at $T=0$
are represented in Figs. 1 and  2.

We apply the local density approximation (LDA) to obtain the local
superfluid density $n_{s}(T,r)$ and the Kosterlitz-Thouless phase
transition temperature $T_{c}(r)$ at each point $r$ of the trap.
The LDA can be used for a potential which is flat on the
characteristic length scale $l$ much greater than the average
distance $R$ between vortices of the normal phase, where $R$ is
estimated as\cite{Kosterlitz}:
\begin{eqnarray}
\label{vortex} \ln\frac{R}{\rho} \sim
\exp \left\{ \left(\ln\frac{T_{c}}{T -
T_{c}}\right)^{1/2}\right\}, \hspace{1cm} {T\rightarrow T_{c}^{+}}.
\end{eqnarray}
This estimation works at temperatures above $T_{c}$, and diverges
at $T=T_{c}$. So we can use the LDA at temperatures sufficiently
close to $T_{c}$, but it breaks down at $T_{c}$. In the LDA
framework, the trapped system has at each point in space exactly
the same properties as an infinite system, using the local density
$n(r)$ instead of the average density $n$ for the infinite system.
Substituting the local density profile Eq.~(\ref{den_pr_ap}) into
the expressions for $n_{s}$ and $T_{c}$ for the infinite
system\cite{Berman_Snoke_Coalson}, we obtain for the local
superfluid density
\begin{eqnarray}
\label{nn55} n_{s} (r) = n (r) -  s\frac{3 \zeta (3) }{2 \pi }
\frac{T^3}{c_{s}^{4}(n(r),Q) M} - s\frac{n(r)
Q}{2Mc_{s}^{2}(n(r),Q)},
\end{eqnarray}
and for the local Kosterlitz-Thouless temperature
\begin{eqnarray}
\label{tct} T_c(r) = \left[\left( 1 +
\sqrt{\frac{32}{27}\left(\frac{M T_{c}^{0}(r)}{\pi
n'(r)}\right)^{3} + 1} \right)^{1/3}   - \left(
\sqrt{\frac{32}{27} \left(\frac{M T_{c}^{0}(r)}{\pi
n'(r)}\right)^{3} + 1} - 1 \right)^{1/3}\right]
\frac{T_{c}^{0}(r)}{ 2^{1/3}}   ,
\end{eqnarray}
where $c_{s}^{2}(r) = \sqrt{\mu^{2}[n(r)] - Q^{2}}/M$ is the local
velocity of sound and $s$ is the spin degeneracy, equal to 4 for
excitons in GaAs
quantum wells. Here
$T_{c}^{0}$ is an auxiliary quantity, equal to the temperature at
which the superfluid density vanishes in the mean-field
approximation (i.e., $n_{s}(T_{c}^{0}) = 0$),
\begin{equation}
\label{tct0} T_c^0 (r) = \left( \frac{2 \pi n'(r) c_s^4 (r) M}{3
\zeta (3)} \right)^{1/3}  .
\end{equation}
In Eqs.~(\ref{tct}) and (\ref{tct0}), $n'(r)$ is
\begin{eqnarray}
\label{nexx} n'(r) = n(r)  - s\frac{n(r) Q}{2Mc_{s}^{2}(r)} .
\end{eqnarray}

For the trapped excitonic system we find the dependence of the
critical minimal total number $N_{c}$ of particles where the
global superfluidity exists on the temperature $T$ at different
$Q$. The total number of particles corresponding to fixed $n_{0}$
is given by
\begin{eqnarray}
\label{Nc} N = 2 \int_{0}^{r_{0}} n(r)d^{2}r ,
\end{eqnarray}
where, again, the exciton density profile $n(r) = n_{0} - \Delta
n(r)$ ($\Delta n(r)$ is given by Eq.~(\ref{den_pr_ap})); $r_{0} =
\mu(2/(\gamma\sqrt{\mu^2 -Q^2}))^{1/2}$ is the root of the
equation $n(\pm r_{0})=0$.

Expressing $n(r)$ in terms of $N$ using Eq.~(\ref{Nc}), and
substituting $n(r)$ in Eq.~(\ref{tct}), we calculate the critical
  total number $N_{c}(T)$ of excitons corresponding to
  Kosterlitz-Thouless phase transition (see Fig.3). Our calculations
show that at fixed temperature
  $T$,  it takes a greater total number of excitons $N_{c}$
to achieve superfluidity as disorder increases.


\section{Trapped indirect magnetoexcitons in coupled quantum wells in
random field}\label{mag}

We now consider indirect excitons in a strong magnetic field
perpendicular to the quantum wells in the presence of the
disorder, extending previous work on a homogeneous gas of excitons
in a magnetic field.\cite{Berman_Snoke_Coalson_magnet} In this
case we neglect
  the transitions between different Landau
levels of the magnetoexciton, including transitions caused by
scattering from the slowly varying spatial confinement  potential
$U(r_{e}, r_{h}) = \frac{1}{2} \gamma(r_{e}^{2} + r_{h}^{2})$ and
the random field potential. We also neglect nondiagonal matrix
elements of the Coulomb interaction between a paired electron and
hole. The region of applicability of these two assumptions is
defined by the inequalities\cite{Ruvinsky_jetp} $\omega_{c} \gg
E_{b}$, $\omega_{c} \gg \sqrt{\left\langle
V_{e(h)}^{2}\right\rangle _{av}}$, where $\omega_{c} =
eH/m_{e-h}$, and $m_{e-h} = m_{e}m_{h}/(m_{e} + m_{h})$ is the
exciton reduced mass in the quantum well plane; $E_{b}$ is the
magnetoexciton binding energy in an ideal ``pure'' system as a
function of magnetic field $H$ and the distance between electron
and hole quantum wells $D$: $E_{b} \sim e^{2}/\epsilon
r_{H}\sqrt{\pi/2}$ at $D \ll r_{H}$ and $E_{b} \sim e^{2}/\epsilon
D$ at $D \gg r_{H}$ ($r_{H} = (eH)^{-1/2}$ is the magnetic
length).\cite{Lerner,Ruvinskiy} Here $\left\langle \ldots
\right\rangle _{av}$ denotes an average over the fluctuations of
the random field. The characteristic length of inhomogeneity is
assumed to be much greater than the magnetic length: $l\gg r_{H}$.
We consider the characteristic length of the random field
potential  $L$ to be much shorter than the average distance
between excitons $r_{s} \sim 1/\sqrt{\pi n}$  ($L \ll 1/\sqrt{\pi
n}$, where $n$ is the total exciton density) similar to
Ref.~[\onlinecite{Berman_Snoke_Coalson}].

It can be shown\cite{Berman_Snoke_Coalson_magnet} that the effective
Hamiltonian
$\hat{H}_{\rm eff}$  of the system of indirect ``dirty''
magnetoexcitons at small
momenta is identical to the Hamiltonian of indirect ``dirty'' excitons without
magnetic field but with magnetic mass $m_{H}$ instead of $M = m_{e} +
m_{h}$. In strong magnetic fields
at $D \gg r_{H}$ the exciton magnetic mass is $m_H \approx
D^{3}\epsilon/(e^{2}r_{H}^{4})$ \cite{Ruvinskiy}. Therefore, for
the trapped system in strong magnetic field we can apply the
expressions for the exciton local density profile, local
superfluid density and local temperature of the Kosterlitz-Thouless
phase transition  for the ``dirty'' trapped (Sec.~\ref{nomag})
system without magnetic field using random field $V_{\rm eff}$, and an
effective external field of confinement $U_{\rm eff}$ instead of $
U_{e}(\mathbf{r}) + U_{h}(\mathbf{r})$, respectively. The
effective  parameter of disorder $Q$ for the system of
magnetoexcitons is given by\cite{Berman_Snoke_Coalson_magnet}
\begin{eqnarray}
\label{Q_res_ap1}  Q (\mathbf{p}, \omega) &=& Q =
\frac{\alpha_{e}^{2}(g_{1} + g_{2}) + \alpha_{h}^{2}(g_{3} +
g_{4})}{64\pi^{4}}m_{H}.
\end{eqnarray}
and the effective field of confinement
has the form
\begin{eqnarray}\label{V_eff}
U_{\rm eff}(\mathbf{R}) &=& \frac{1}{\pi r_{H}^{2}} \int \exp\left(
-\frac{(\mathbf{R}- \mathbf{r})^{2}}{r_{H}^{2}}\right) \left[
U_{e}(\mathbf{r}) + U_{h}(\mathbf{r}) \right] d \mathbf{r}.
\end{eqnarray}
Eq.~(\ref{V_eff}) is valid if the characteristic length $l$ of
inhomogeneity of the trapping potential
$U(\mathbf{r}_{e},\mathbf{r}_{h})$ is much greater than the
magnetoexciton mean size $r_{exc} \approx r_{H}$. In a strong
magnetic field we obtain $U_{\rm eff}(\mathbf{R}) = U(\mathbf{R})
= \gamma R^{2}/2$. Therefore, magnetic field does not change the
effective trapping potential.

Since magnetic field affects the effective Hamiltonian only by
replacing the excitonic mass $M$ by effective the magnetic mass
$m_{H}$,\cite{Berman_Snoke_Coalson_magnet} the magnetoexciton
local  density profile can be obtained by substitution the
effective disorder parameter $Q$ (Eq.~(\ref{Q_res_ap1})) into
Eq.~(\ref{den_pr_ap}) and effective magnetic mass $m_{H}$ instead
of $M$ and $s=1$ into the expression for $\mu$ (Eq.~(\ref{Mu}))
and Eq.~(\ref{den_pr_ap}). As we can see from this substitution,
the magnetoexciton local density at a fixed local point $\mathbf{r}$
decreases as magnetic field increases proportional to the magnetic
mass $m_{H} \sim H^{2}$ when $D \gg r_{H}$. The negative correction
to the local exciton density is also an increasing function of the
interwell separation $D$ because $m_{H} \sim D^{3}$. The local
superfluid density and the local Kosterlitz-Thouless phase transition
temperature can be obtained by replacing $M$ by $m_{H}$ and
substitution of the magnetoexciton local density profile in
Eqs.~(\ref{nn55}), ~(\ref{tct}),~(\ref{tct0}) and~({\ref{nexx}).
As we can see from Eq.~(\ref{Q_res_ap1}), increasing magnetic
field $H$, corresponding to higher $Q$, results in higher minimal
total number $N_c$ required to observe superfluidity at fixed $T$.


\section{Discussion}
\label{discussion}

As we can see from Eq.~(\ref{tct}), the critical temperature $T_c$
for superfluidity at $r=0$ is the same as that of an infinite
system with $n = n_0$.  Since at $r > 0$ the local  exciton
density decreases, as  we go farther from the minimum, the
Kosterlitz-Thouless critical temperature and superfluid density
will decrease. In general, there will be a superfluid centered at
$r=0$ and a normal phase outside this area; the size of the
superfluid region will shrink as $T$ increases.

The disorder is found to deplete the superfluid, making the
superfluid region smaller than it would be in the absence of
disorder, according to Eq.~(\ref{den_pr_ap}) (see Figs.~1 and 2).
A ``clean'' system is always superfluid at $T=0$, but a ``dirty''
system will have a quantum phase transition to the superfluid
state at some critical density (see Fig.~3) even at $T=0$.

In a strong
magnetic field
$H$, the correction of the local density is found to increase with
magnetic field as $H^{2}$ and with interwell separation $D$ as
$D^{3}$. As magnetic
field $H$ increases,
the minimum total number $N_c$ required to observe superfluidity
increases at fixed
  $T$. The critical effective disorder parameter increases with magnetic
  field as $H^{2}$ at $T=0$.

\begin{center}
{\bf
Acknowledgements}
\end{center}

O. L.~B. wishes to thank the
participants of the First
International Conference on Spontaneous
Coherence in Excitonic
Systems (ICSCE) in Seven Springs PA for many
useful and
stimulating discussions. Yu. E.~L. was supported by the
INTAS
grant. D. W.~S. and R. D.~C. have been supported by the National
Science Foundation.


\newpage

\begin{figure}
\rotatebox{270}{
\includegraphics[width = 16cm, height = 17cm]{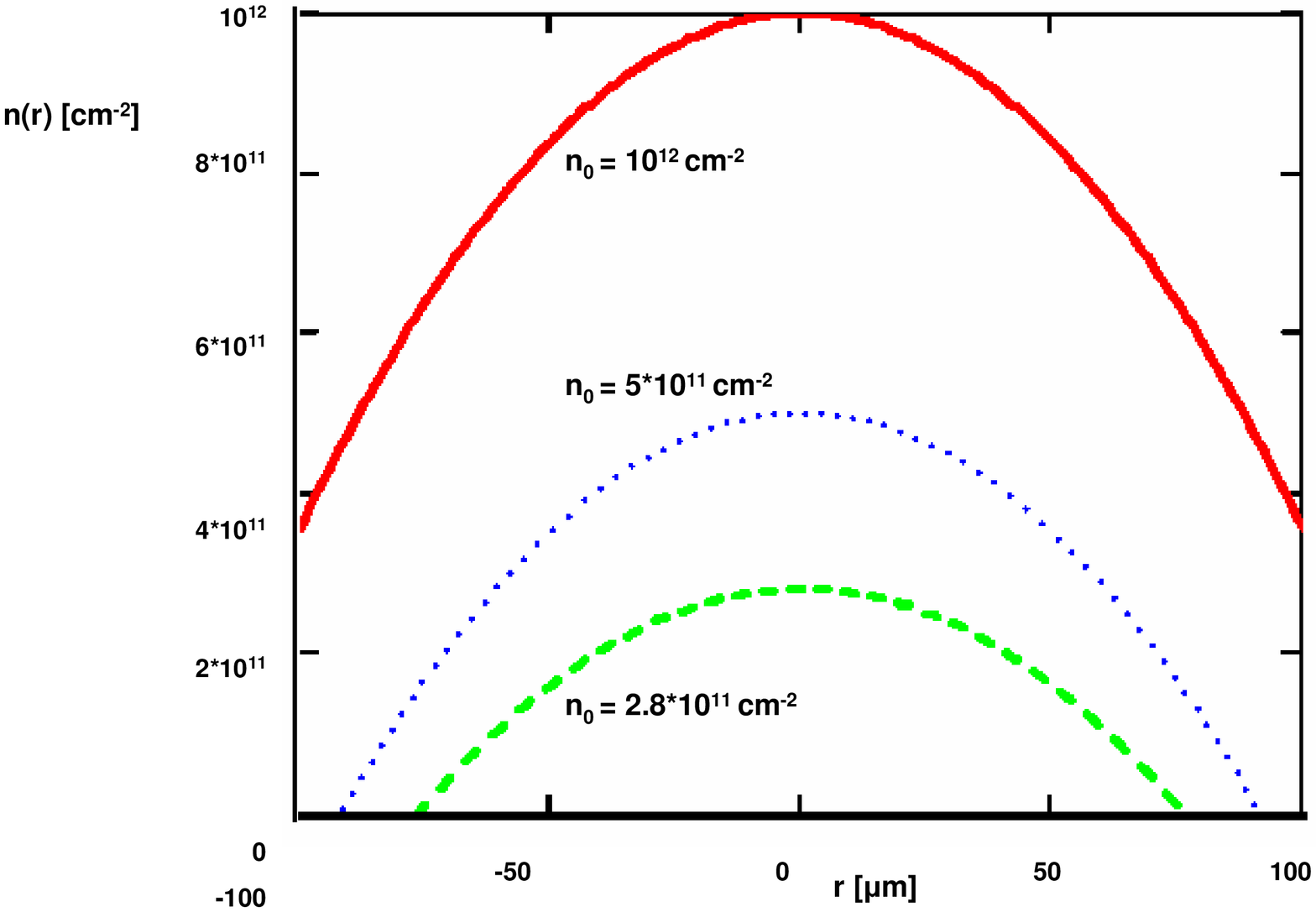}}
\caption{Exciton density profile at $T=0$ for different $n_{0}$ at
the disorder parameter $Q = 0.58 meV$; interwell distance $D = 14
nm$;  $\gamma  = 40 eV/cm^{2}$; dielectric constant $\epsilon =
13$. The parameters of the system are taken from
Ref.[\onlinecite{Snoke-apl}].}
\end{figure}

\newpage

\begin{figure}
\rotatebox{90}{
\includegraphics[width = 16cm, height = 17cm]{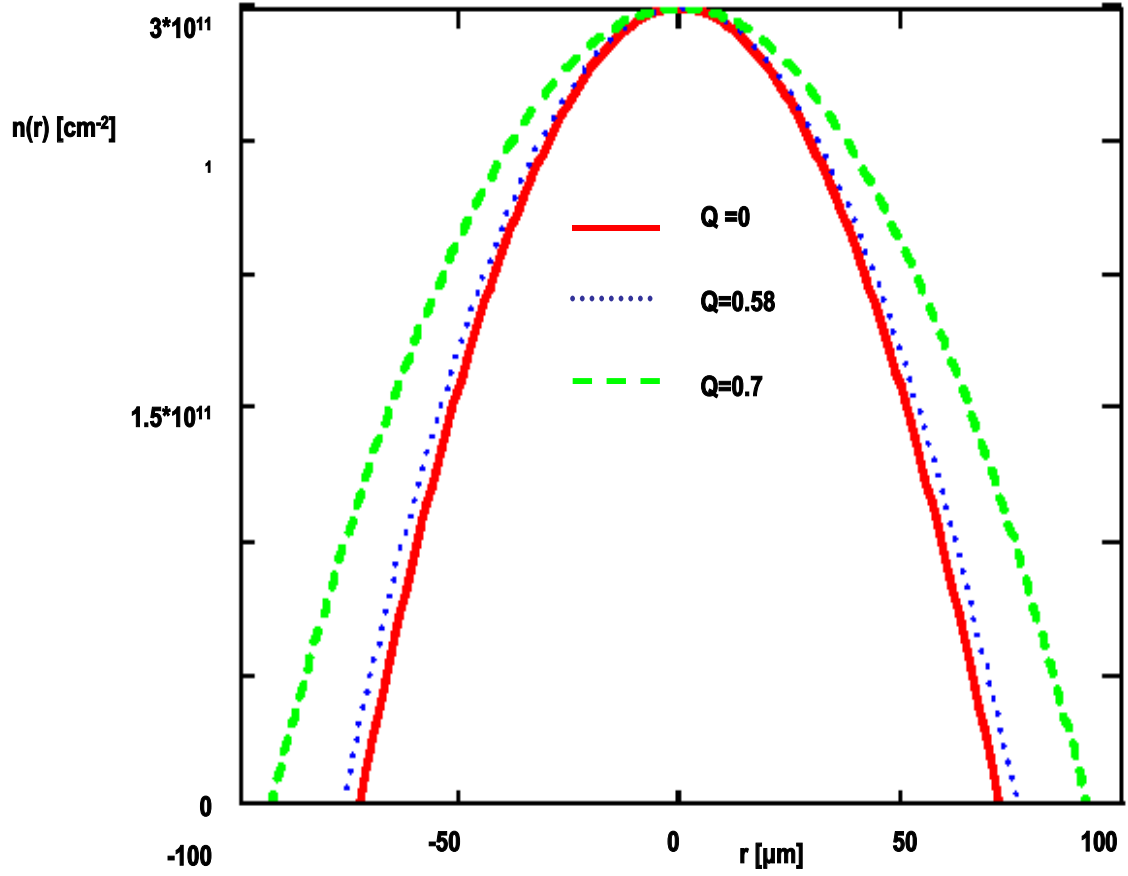}}
\caption{ Exciton density profile at $T=0$ for different disorder
parameters $Q$ (in units of $meV$) at $n_{0} = 3\times 10^{11}
cm^{-2}$; interwell distance $D = 14 nm$;  $\gamma  = 40
eV/cm^{2}$; dielectric constant $\epsilon = 13$.The parameters of
the system are taken from Ref.[\onlinecite{Snoke-apl}].}
\end{figure}

\newpage

\begin{figure}
\rotatebox{90}{
\includegraphics[width = 16cm, height = 17cm]{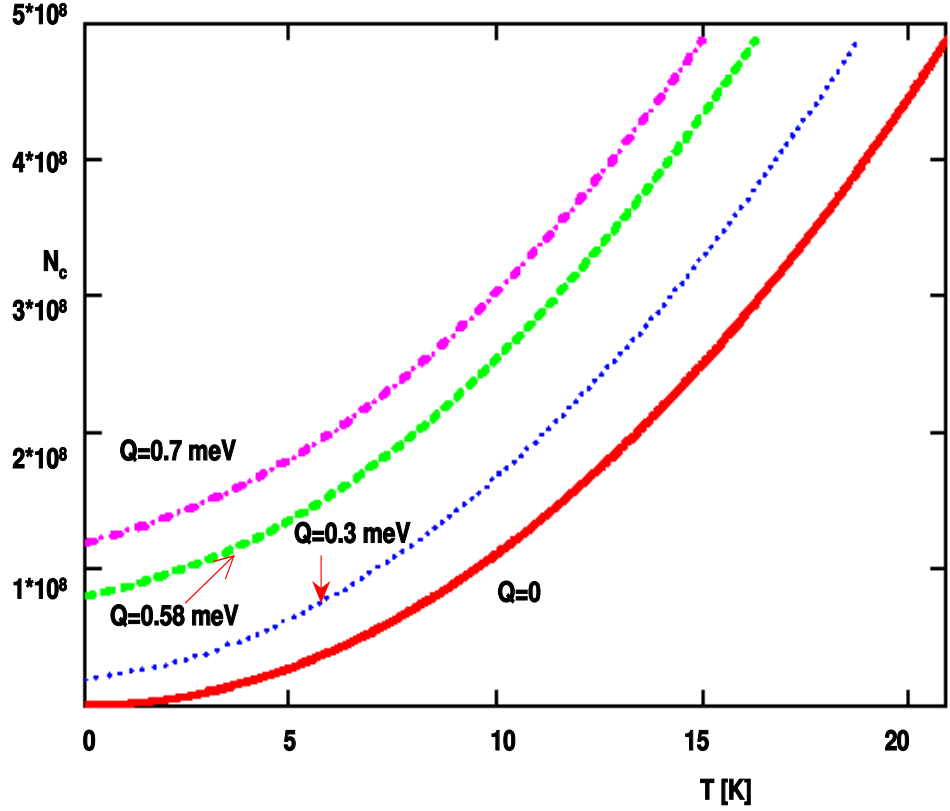}}
\caption{Critical total number of particles $N_{c}(T)$
corresponding to the Kosterlitz-Thouless phase transition as a
function of temperature $T [K]$ for different disorder parameters
$Q [meV]$ at interwell distance $D = 14 nm$;  $\gamma  = 40
eV/cm^{2}$; dielectric constant $\epsilon = 13$.The parameters of
the system are taken from Ref.[\onlinecite{Snoke-apl}].}
\end{figure}

\newpage

\end{document}